\documentclass{elsart}

\usepackage{graphicx}
\usepackage{amssymb}

\begin{document}

\begin{frontmatter}

\title{Theory of Nucleosome Corkscrew Sliding in the Presence of Synthetic DNA Ligands}

\author[label1]{Farshid Mohammad-Rafiee}
\ead{farshidm@iasbs.ac.ir}
\author[label2]{Igor M. Kuli\'{c}}
\ead{kulic@mpip-mainz.mpg.de}
\author[label2]{Helmut Schiessel}
\ead{heli@mpip-mainz.mpg.de}
\address[label1]{Institute for Advanced Studies in Basic Sciences, Zanjan 45195-159, Iran}
\address[label2]{Max-Planck-Institut f\"{u}r Polymerforschung, Theory Group, POBox 3148, D 55021 Mainz, Germany}

\begin{abstract}
Histone octamers show a heat-induced mobility along DNA. Recent
theoretical studies have established two mechanisms that are
qualitatively and quantitatively compatible with {\it in vitro}
experiments on nucleosome sliding: Octamer repositiong through
one-basepair twist defects and through ten-basepair bulge defects.
A recent experiment demonstrated that the repositioning is
strongly suppressed in the presence of minor-groove binding DNA
ligands. In the present study we give a quantitative theory for
nucleosome repositioning in the presence of such ligands. We show
that the experimentally observed octamer mobilities are consistent
with the picture of bound ligands blocking the passage of twist
defects through the nucleosome. This strongly supports the model
of twist defects inducing a corkscrew motion of the nucleosome as
the underlying mechanism of nucleosome sliding. We provide a
theoretical estimate of the nucleosomal mobility without
adjustable parameters, as a function of ligand concentration,
binding affinity, binding site orientiation, temperature and DNA
anisotropy. Having this mobility at hand we speculate about the
interaction between a nucleosome and a transcribing RNA polymerase
and suggest a novel mechanism that might account for polymerase
induced nucleosome repositioning.
\end{abstract}

\end{frontmatter}

\section{Introduction}
\label{section1}

The dynamics of folding and unfolding of DNA within the chromatin
complex is of vital importance for the regulation of genes. The
basic unit of chromatin is the nucleosome where DNA is wound in 1
and 3/4 lefthanded superhelical turns around an octamer of histone
proteins \cite{luger97}. Roughly $75\%$ of all eucaryotic DNA is
tightly associated to such protein spools. All that
intranucleosomal DNA is usually not accessible to DNA binding
proteins \cite{workman98}, leading to the puzzling question of how
these proteins can find their hidden target sites. And even more
surprising is the fact that most m-RNA coding genes are also
covered with tens to hundreds of nucleosomes. So how does a
transcribing RNA polymerase deal with all the octamers that it
encounters on its way? Can it "get around" the nucleosomes or have
the nucleosomes to be removed before transcription is possible?

An important insight in that respect is the fact that nucleosomes
are highly dynamical objects. It has been demonstrated through
competitive protein binding that thermal fluctuations induce
spontaneous unwrapping of nucleosomal DNA at the ends of its
wrapped portion \cite{polach95,anderson00}. This leads to a
transient opportunity for proteins to bind to nucleosomal DNA.
Another important mechanism is nucleosome "sliding". It has been
observed under well-defined {\it in vitro} conditions that
nucleosomes spontaneously reposition themselves along DNA
\cite{pennings91,meersseman92,pennings94,flaus98} transforming
nucleosomal DNA into free DNA and {\it vice versa}. Heat induced
repositioning is, however, a slow process happening on time scales
of minutes to hours. The {\it in vivo} octamer repositioning has
thus to be catalyzed by ATP consuming machines, so-called
chromatin remodelling complexes \cite{kornberg99,becker02}.

Repositioning experiments (reviewed in detail in
Ref.~\cite{schiessel03}) are typically performed on short pieces
of DNA of length 200 to 400 basepairs (bp) that contain one or two
positioning sequences. Repositioning is detected with the help of
2D gel electrophoresis making use of the fact that complexes with
octamers close to one of the DNA termini show a higher
electrophoretic mobility \cite{pennings91,meersseman92,pennings94}
than complexes where the octamer is associated to the center of
the DNA fragment. Another approach \cite{flaus98} uses a
chemically modified histone protein that induces a cut on the
nucleosomal DNA. The general outcome of these studies is as
follows: (1) Heat induced repositioning is a slow process taking
place on the time scale of minutes to hours
\cite{pennings91,flaus98} at elevated temperatures (say
$37^{\circ}$) but it is strongly suppressed at lower temperatures
(say $5^{\circ}$). (2) The octamer is found at a preferred
position (as mentioned above the DNA contains a positioning
sequence!) or multiples of 10 bp (the DNA helical pitch) apart
\cite{pennings91,flaus98}. (3) There is a preference for end
positions \cite{pennings91}. (4) For longer DNA segments there is
no evidence for a long-range repositioning \cite{meersseman92}.
(5) In the presence of linker histones (H1 or H5) nucleosome
mobility is suppressed \cite{pennings94}.

What is the origin of the nucleosome mobility? An ordinary sliding
of the DNA on the protein spool is energetically too costly: The
interaction between the DNA and the octamer is localized at 14
binding sites where the minor groove of the DNA faces the octamer
surface \cite{luger97}, each contributing roughly $6 k_B T$ pure
adsorption energy \cite{schiessel03} ($k_B T$: thermal energy). A
bulk sliding motion would involve the simultaneous breakage of
these 14 point contacts, an event that certainly would never occur
spontaneously. A rolling motion of the octamer along the DNA is
also not possible: Due to the helical wrapping path the cylinder
would simply roll off the DNA.

Repositioning must involve intermediate states that have a lower
energetic barrier. Two commonly accepted possible mechanisms
\cite{schiessel03,flaus03} are based on small defects that
spontaneously form in the wrapped DNA portion and propagate
through the nucleosome: 10 bp bulges \cite{schiessel01,kulic03a}
and 1 bp twist defect \cite{kulic03b}. The basic idea of bulge
defects is as follows: As a first step the DNA unpeels
spontaneously from one of the termini of the wrapped portion
\cite{polach95,anderson00}. Subsequently some DNA is pulled in
before the chain readsorbs creating an intranucleosomal DNA bulge
that carries some extra length $\Delta L$. Once a loop has formed
it diffuses along the wrapped DNA portion and finally leaves the
nucleosome at either end. If the loop happens to come out at the
end where it has been created nothing happens. But if the loop
leaves at the other end the extra length $\Delta L$ has been
transported through the nucleosome and the octamer is repositioned
by $\Delta L$ along the DNA. A careful quantitative analysis
\cite{kulic03a} showed that the cheapest small loop that can be
formed has a length of 10 bp. Such a loop is not twisted; the next
planar loop, a 20 bp bulge, is much more expensive. However, even
the creation of a 10 bp loop is very costly: Its formation
requires about $20 k_B T$ desorption and bending energy and thus
constitutes a very rare event \cite{schiessel03}. As a consequence
the corresponding diffusion constant of the octamer along the DNA
is very small, namely on the order of $D \approx 10^{-16}cm^2/s$.
Thus typical repositioning times on a 200 bp DNA fragment are on
the order of an hour, in reasonable agreement with the
experimental data \cite{pennings91,flaus98}. The strong
temperature dependence and most strikingly the preference for 10
bp steps -- corresponding to the extra length stored in the
cheapest loops -- is also in excellent agreement with the
experiments. So at first sight it seems that the loops are in
every respect a promising candidate for the mechanism underlying
repositioning. There is, however, one serious caveat: We found
that larger loops beyond one persistence length of DNA (roughly
150 bp) are easier to form than 10 bp bulges since such loops show
a small curvature and have less desorbed binding sites
\cite{kulic03a}. Of course, for short DNA segments such loops
cannot occur. However, experiments with DNA segments of length
$\approx 400$ bp have also not shown any signature of a long range
nucleosome repositioning \cite{meersseman92}.

We therefore reconsidered the underlying mechanism and checked
whether the experimental observations would also be consistent
with repositioning via twist defects \cite{kulic03b}. The basic
idea of repositioning via twist defects is that thermally
activated defects form spontaneously at the termini of the wrapped
DNA portion. There are two types of twist defects: (a) defects
with a missing bp requiring the DNA to stretch and overtwist
between its two neighboring nucleosomal binding sites and (b)
defects with an extra bp thus leading to\ a compressed and
undertwisted piece of DNA between two nucleosomal point contacts.
As in the case of bulges a twist defect might diffuse around a
nucleosome releasing its stored length (here $\pm 1$ bp) at the
other end. The result of such an event is that the octamer makes a
step by one bp {\it and} a rotation by $36^{\circ}$ around the DNA
axis; or {\it vice versa} one might say the DNA performs a
corresponding corkscrew motion on the nucleosome. The cost of
forming a one bp twist defect was estimated to be on the order of
$9 k_B T$ \cite{kulic03b}. The shorter defect length involved in
twist defects as compared to bulges, is thus dramatically
overcompensated by their lower activation cost. In fact, we
estimated that twist defects lead to a nucleosome diffusion
constant on the order of $D \approx 10^{-12}cm^2/s$ that is 4
orders of magnitude larger than the one predicted by loop defects.
The typical repositioning times on a 200 bp piece of DNA are thus
predicted to be on the order of a second, a time much shorter than
in the experiments! Even worse, the predicted dependence of the
dynamics on temperature is much too weak and there is no
"built-in" mechanism for 10 bp steps of the octamer. The
experimentally observed preference for positions 10 bp apart
manifesting itself in characteristic bands in the products of a
gel electrophoresis \cite{pennings91,meersseman92} seems to be
inconsistent with this mechanism -- at least at first sight.

Here comes into a play an important additional feature of the
repositioning experiments, namely that they are typically
performed with DNA segments containing positioning sequences,
especially the sea urchin 5S positioning element
\cite{pennings91,meersseman92,pennings94}. The characteristic
feature of this sequence is that it shows a highly anisotropic
bendability of the DNA. If the underlying mechanism of
repositioning is a 1 bp twist defect, then the DNA has to bend in
the course of a 10 bp shift in all directions, and thus has to go
over a barrier. In the case of the standard "5S-RNA" this barrier
is on the order of $9-10 k_B T$ \cite{anselmi00,mattei02}. The
typical repositioning times on a 200 bp DNA segment are now 2 to 3
orders of magnitude longer, i.e., they are on the order of an hour
as in the loop case! Now it is a simple matter of equilibrium
thermodynamics that the probability of finding the DNA wrapped in
its preferred bending direction is much higher than in an
unfavorable one. This means, however, that also in the case of 1
bp defects one would find nucleosomes mostly at the optimal
position or 10, 20, 30 etc bp apart, i.e., at locations where
still most of the positioning sequence is associated with the
octamer and this in the preferred bending direction. The bands in
the gel electrophoresis experiments have then to be interpreted as
reflecting the Boltzmann distribution of the nucleosome positions.
In other words, both the 10 bp bulge and the 1 bp twist defect
lead in the presence of a rotational positioning sequences to
pretty much the same prediction for the experimentally observed
repositioning, even though the elementary motion is fundamentally
different!

There are many ways to design experiments that could help to
identify the mechanism that underlies nucleosomal mobility. The
most obvious idea is to use a DNA template with less exotic
mechanical properties (in fact, the 5S positioning element is the
strongest natural positioning sequence known so far). If
nucleosomes move via loop defects a more isotropically bendable
DNA should not speed up the dynamics whereas it would have a
strong impact on the corkscrew mechanism. In fact, the experiment
by Flaus and Richmond \cite{flaus98} goes in that direction. They
used a DNA fragment of length 438 bp that featured two positioning
sequences where two nucleosomes assembled, each at a unique
position. These positions were also found when mononucleosomes
were assembled on shorter fragments that included only one of the
two positioning elements. The authors studied the degree of
repositioning of the mononucleosomes on such shorter fragments
(namely nucleosome A on a 242 bp- and nucleosome B on a 219
bp-fragment) as a function of heating time and temperature. It was
found that the repositioning rates increase strongly with
temperature but also depend on the positioning sequence (and/or
length of the fragment). The difference of repositioning for the
two sequences is remarkable: at 37${{}^{\circ }}$C one has to wait
$\sim 90$ minutes for the A242 and more than 30 hours for the B219
for having half of the material repositioned. For nucleosome B
which showed a slower repositioning the set of new positions were
all multiples of 10 bp apart (namely at a 20, 30, 40, 50
bp-distance from the starting position), i.e., they all had the
same rotational phase. On the other hand, nucleosome A did not
show such a clear preference for the rotational positioning. It
was argued that these differences reflect specific features of the
underlying base pair sequences involved. Nucleosome B is complexed
with a DNA sequence that has AA/AT/TA/TT dinucleotides that show a
10 bp periodicity inducing a bend on the DNA whereas nucleosome A
is positioned via homonucleotide tracts. These observations are
clearly consistent with the twist defect picture where the
corkscrew motion of nucleosome B is suppressed by the
anisotropically bendable DNA template.

Another experimental approach was taken recently by Gottesfeld
{\it et al.} \cite{gottesfeld02}. The authors considered a 216 bp
DNA fragment that again contained the sea urchin 5S rDNA
nucleosome positioning sequence. They also followed the heat
induced nucleosome repositioning but this time in the presence of
pyrrole-imidazole polyamides, synthetic minor-groove binding DNA
ligands that are designed to bind to specific target sequences.
Experiments have been performed in the presence of one of 4
different ligands, each having one binding site on the nucleosomal
DNA. The general outcome of this study was as follows: (1) A
one-hour incubation at $37^\circ$ in the absence of any ligand
leads to a redistribution of the nucleosomes. (2) In the presence
of 100 nM ligands {\it no} repositioning of nucleosomes is
detected after such an incubation if the target sequence of this
specific ligand faces the solution when the DNA is bent in its
preferred direction. (3) If a ligand has been added whose binding
site faces the octamer in its preferred rotational frame, the
ligand has no detectable effect on the reposition dynamics.

This raises the question whether this experiment is capable to
distinguish between loop- and twist-defect induced nucleosome
mobility. Since the ligands bind into the minor groove (cf. the
co-crystal complexes between nucleosomes and such ligands
\cite{suto03}) it is quite likely that a bound ligand will block
the overall corkscrew motion of the DNA; the DNA can only rotate
on the nucleosome up to a point where the bound ligand comes close
to one of the 14 binding sites. A further rotation of the DNA is
not possible because of steric hindrance and twist defects are
reflected once they encounter the ligand site. In other words: The
observed suppression of mobility through ligand binding agrees
qualitatively well with the twist defect picture. What about a
bulge defect encountering a bound ligand? In this case the answer
is not obvious. In a first approximation one should expect that a
bound ligand does not hinder bulge diffusion -- at least
sterically. Of course, the ligand might locally alter the DNA
elastic properties so that we cannot give here a definite answer.
But obviously the influence of ligand binding on nucleosome
mobility supports much more the idea of twist diffusion as the
underlying mechanism. The aim of this study is to demonstrate that
the twist diffusion picture is indeed compatible with the
experimental data presented in Ref.~\cite{gottesfeld02}

In the next section we provide a theoretical model for nucleosome
repositioning in the presence of DNA ligands. We make use of our
recent results on repositioning via twist defect in the absence of
ligands \cite{kulic03b} that essentially provides us with the
nucleosomal diffusion constant as a function of temperature and
underlying DNA sequence. Assuming thermodynamic equilibrium we
will then calculate the diffusion constant in the presence of
ligands. We find -- in agreement with the experiments -- that in
the presence of 100nM ligands the repositioning on the 5S
positioning sequence is essentially completely blocked if the
ligand binding site prefers to face the solution. On the other
hand, when the binding site faces the octamer, the ligands have a
negligible influence on the nucleosome mobility.

Knowing the nucleosome mobility in the various cases we speculate
in Section 3 what happens when a transcribing RNA polymerase
encounters a nucleosome. In fact, this situation has been
investigated by Gottesfeld {\it et al.} \cite{gottesfeld02} in
their study with synthetic ligands. We suggest that in some of the
cases the RNA polymerase pushes the nucleosome in front of it in a
corkscrew fashion. When the terminus of the DNA template is
reached the nucleosome becomes "undressed" and the other end of
the DNA might bind to the exposed binding sites. As a result the
RNA polymerase seems to have gotten around the nucleosome with the
octamer having been transferred to a location upstream. Our model
gives thus an alternative explanation to the popular idea of RNA
polymerase transcribing through a nucleosome in a loop
\cite{studitsky94,studitsky95,studitsky97,bednar99}.

\section{Autonomous Repositioning}
\label{section2}

\subsection{Nucleosome Mobility in the Absence of Ligands}

Let us first consider the repositioning of nucleosome along DNA
induced via 1 bp twist defects in the absence of ligands. This has
been recently studied theoretically in detail \cite{kulic03b};
here we will restrict ourselves to a short presentation of the
results. The basic idea is that a twist defect might form
spontaneously at either end of the wrapped DNA portion. Such a
defect can carry a missing or an extra bp. A defect is typically
localized between two neighboring nucleosomal binding sites, i.e.,
it is localized within one helical pitch, 10 bp. This short
portion of DNA is stretched (compressed) and overtwisted
(undertwisted). The energy of a $\pm 1$ bp twist defects was
estimated from the combined stretch and twist moduli of DNA
including the (here unfavorable) twist-stretch coupling to be on
the order of $9 k_B T$ \cite{kulic03b}. That means that one finds
for a given time a twist defect only on one of around thousand
nucleosomes.

Once a twist defect has formed, it can diffuse through the wrapped
DNA portion. The nucleosome provides in total 13 positions for the
defect between the 14 binding sites. A defect (say a "hole" with a
missing bp) can move from one position to the next in the fashion
of an earthworm creep motion: The bp that is in contact with a
binding site moves towards the defect, leading to an intermediate
state where the defect is stretched out over 20 bp, releasing
elastic energy but paying desorption energy. When the next bp
binds to the nucleosome the twist defect has moved to the
neighboring location. During this process the kink goes over a
barrier; its energy was estimated from the adsorption energy per
contact and the DNA elasticity to be on the order of $2 k_B T$
\cite{kulic03b}. Of course, not all twist defects that have formed
will reach the other end of the nucleosome, most fall off at the
terminus at which they have been created. In fact, one can show --
assuming that all 13 possible defect locations are energetically
equivalent -- that only $1/13$ of the defects are "successful",
only this fraction contributing to the nucleosomal mobility.

Putting all these points together we were able to estimate the
diffusion constant of the nucleosome along DNA to be $D_0 \approx
580bp^2/s\approx7 \times 10^{-13}cm^2/s$. As mentioned in the
introduction this is surprisingly fast -- especially much faster
than the experimental values
\cite{pennings91,meersseman92,pennings94,flaus98}. We explained
this discrepancy by the mechanical properties of the underlying
DNA template. A nucleosome performing a (random) corkscrew motion
might encounter a very bumpy energy landscape. Especially,
positioning sequences like the commonly used 5S sequence show an
anisotropic bendability. In that case the elastic energy of the
bent DNA is a periodic function of the nucleosome position with
the helical pitch being the period. We approximated this energy by
an idealized potential of the form $U(l)=\left(A/2\right)\cos(2\pi
l/10)$ with $l$ being the number of the bp say at the dyad axis
and $A$ denoting the difference in elastic energy between the
optimal and the worst rotational setting. Of course, these
oscillations die out completely when the nucleosome leaves the
positioning sequence, i.e., if it has moved around 140 bp. Since
the templates are usually quite short (for instance, 216 bp
\cite{gottesfeld02}) the nucleosome will always feel the
rotational signal from the positioning sequence so that our
elastic energy should provide a reasonable description. In that
case the nucleosomal diffusion constant is reduced as follows
\cite{kulic03b}:
\begin{equation}
D=\frac{D_0}{I_0^2\left(A/2k_B T\right)}\simeq \left\{
\begin{array}{ll}
\frac{D_0}{1+A^2/8\left(k_B T\right)^2} & \mbox{for}\;A < k_B T \\
D_0 \frac{\pi A}{k_B T} e^{-A/k_B T} & \mbox{for}\;A
\gg k_B T%
\end{array}
\right.  \label{deff}
\end{equation}
with $I_0$ being the modified Bessel function and $D_0$ denoting
the diffusion constant for homogenously bendable DNA, $D_0 \approx
580bp^2/s$. For the sea urchin 5S positioning element one has
$A\approx 9 k_B T$ \cite{anselmi00,mattei02} leading to a reduced
mobility with $D \approx 2\times 10^{-15}cm^2/s$.

\subsection{Nucleosome-Ligand Cocomplex: Equilibrium Properities}

In this subsection we determine the equilibrium properties of a
nucleosome in the presence of a finite concentration $[L]$ of a
synthetic ligand targeting one specific site on the nucleosomal
DNA. In Fig.~1 we represent the different possible states by nodes
and the possible pathways from one state to the next state by
connecting lines. Fig.~1(a) shows the case of a DNA template with
an isotropic bendability. The upper row of symbols represent
nucleosomes at different positions without the ligand being bound.
Each circle with a hole (state "1") represents a state where a
ligand can bind, i.e., states where the ligand binding site
(assumed to be located on the wrapped DNA portion) faces away from
the octamer. In this case a ligand can bind leading to a state
that is represented by an open circle, the nucleosome, with a
"bound" black circle, the ligand (state "0"). We assume that in
this case the nucleosome looses its mobility, i.e., we have no
line connecting this state to a neighboring state. Before the
nucleosome can "slide" to a neighboring position the ligand has to
unbind, i.e., one has to go back to state "1". If the nucleosome
is in a position where the ligand binding site faces the octamer
(open circle, state "2") the site is blocked. At these positions
the nucleosome mobility is not affected by the ligands. For
simplicity, we will assume here that always 5 consecutive bp
positions (correspond to one half turn of the corkscrew motion)
have the ligand binding site exposed to the solvent and represent
these 5 position by {\it one} circle with hole. Likewise the other
5 positions have been lumped together into the open circle.

\begin{figure}
\begin{center}
\includegraphics*[width=11cm]{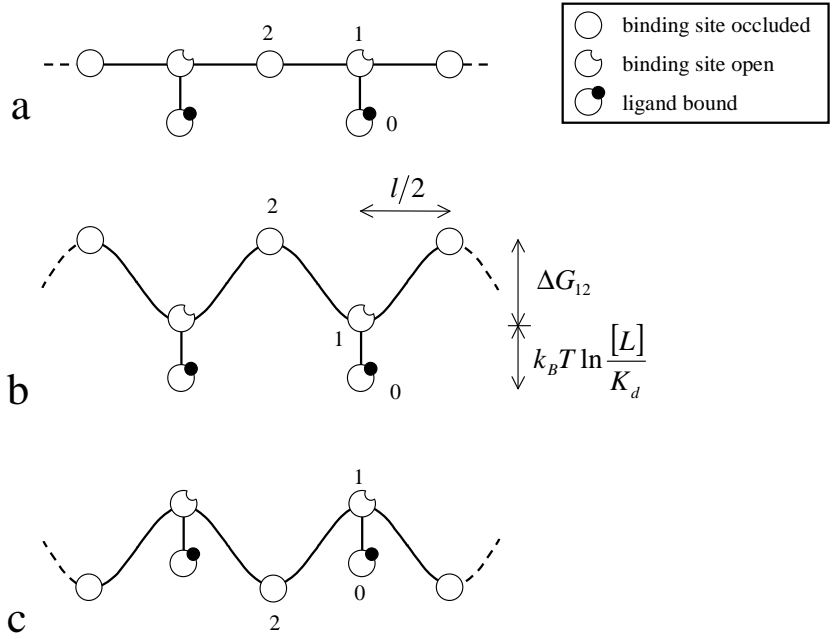}
\end{center}
\caption{Nucleosome repositioning in the presence of ligands. (a)
In the case of a homogenously bendable DNA template the states
with exposed binding sites ("1") and occluded ones ("2") have the
same elastic energy, $\Delta G_{12}=0$. State "0" represents the
immobile state with a bound ligand. (b) and (c) For templates with
a rotational positioning sequence state "1" and "2" have different
elastic energies. In case (b) the preferred rotational frame with
respect to the octamer shows an open binding site, $\Delta
G_{12}>0$, in case (c) a closed one, $\Delta G_{12}<0$. Each node
in this scheme represents 5 consecutive bp positions so that the
periodicity of the bending potential corresponds to one helical
pitch.}
\end{figure}

Fig.~1(b) and (c) show the case of a rotational positioning
sequence as used in the experiment \cite{gottesfeld02}. In case
(b) the situation is such that the ligand can bind when the DNA
sits on the nucleosome in its preferred bending direction. In this
case the states "1" sit in the potential wells of the elastic
energy landscape. We call the difference in the DNA bending energy
between the top and the bottom $\Delta G_{12}=G_2 -G_1$. The case
$\Delta G_{12}>0$ correspond to the situation where a ligand can
most effectively bind to the nucleosome and block the
repositioning (in the experiment this corresponds to the ligands 1
and 4 \cite{gottesfeld02}). Fig.~1(c) depicts the other extreme
where the binding site faces the octamer in the preferred
rotational frame (this corresponds to ligands 2 and 3 in the
experiment \cite{gottesfeld02}).

We denote by $p_i$ the probability for the nucleosome to be in
state $i$, $i=0,1,2$. Detailed balance relates these probabilities
as follows:

\begin{equation}
\frac{p_{2}}{p_{1}}=\frac{\omega _{12}}{\omega _{21}}=e^{-\Delta
G_{12}/k_B T}=f \label{db1}
\end{equation}
and
\begin{equation}
\frac{p_{0}}{p_{1}}=\frac{\omega _{10}}{\omega
_{01}}=K=\frac{[L]}{K_d} \label{db2}
\end{equation}
Here $\omega _{ij}$ denotes the transition rate from state $i$ to
state $j$, $f$ the Boltzmann weight between state 1 and 2, $K$ the
equilibrium constant for the ligand and $K_d$ its dissociation
constant. Eqs.~\ref{db1} and \ref{db2} together with
$p_{0}+p_{1}+p_{2}=1$ yield immediately the occupation
probabilities for the three states:
\begin{equation}
p_{0}=\frac{K}{1+K+f}  \label{p0}
\end{equation}
\begin{equation}
p_{1}=\frac{1}{1+K+f}  \label{p1}
\end{equation}
and
\begin{equation}
p_{2}=\frac{f}{1+K+f}  \label{p2}
\end{equation}

Let us consider some special cases. In the absence of ligands one
has $[L]=0$ and $K=0$. For a homogeneous template one finds in
addition the Boltzmann weight $f$ (defined in Eq.~\ref{db1}) to
equal unity. Then $p_1=p_2=1/2$ and, of course, $p_0=0$. Using a
rotational positioning sequence the nucleosome prefers to be in
its optimal rotational frame. Then for $K=0$ one finds
$p_2/p_1=f$. For the 5S positioning element one has $f\approx
e^{-9}$ so that state "1" is populated with a roughly 10000 times
higher probability then state "2". This might explain the band
structure with a 10 bp periodicity as observed in most
repositioning experiments \cite{pennings91,flaus98}. The presence
of ligands changes the relative weight of the different states.
Consider, for instance, a rotational positioning sequence with the
ligand binding site facing inwards for the preferred rotational
frame ($\Delta G_{12}<0$ and hence $f>1$, cf. Fig.~1(c)). The
relative weight of finding the nucleosome at its mechanically
unfavorable position, state "0" and "1", will then increase in the
presence of ligands. This leads to the intriguing possibility that
the nucleosome changes its preferred position, i.e.,
$p_0+p_1>p_2$. This is in fact the case if $K+1>f$. This means
that for sufficiently high concentration and affinity ligands can
in principle overrule a positioning sequence! The ligands used in
the experiment \cite{gottesfeld02} have dissociation constants
ranging from $0.7$ to $6.0 nM$. For a strong positioning sequence
$f$ is much too large for the above inequality to hold for ligand
concentration typically used in the experiment, say $100nM$.
However, for less strong sequences this might play a role. Also in
the case of a 146 bp template corresponding to the total wrapping
length as consider by Suto {\it et al.} \cite{suto03} it might
well be that ligands shift the preferred centered DNA position to
an off-centered position. A 1-9 bp shift would cost the opening of
one binding site but might allow ligands to bind more effectively,
especially if the binding site(s) at the centered DNA positions
are (partially) occluded. In one case \cite{suto03} (polyamid 2)
such an effect might have been indeed observed, cf. Fig.~5 in that
paper.

\subsection{Nucleosome Mobility in the Presence of Ligands}

We are now in the position to determine the diffusion constant of
a nucleosome along DNA in the various cases. The diffusion
constant can be determined from the average of the diffusion
constant for the nucleosome to jump from state "1" to one of the
two neighboring states "1" and that of going from "2" to
neighboring "2's". Let us denote by $\omega_1$ the rate to go from
a given state "1" to the next position "1" to the right and by
$\omega_2$ the rate of jumps to the right from "2" to "2". Then
the diffusion constant is given by
\begin{equation}
D=\left(p_1\omega_1+p_2 \omega_2\right)l^2
 \label{diffusion}
\end{equation} where $l$ is the jump length, here $l=10 bp$.
Using Kramers' rate theory \cite{haenggi90} it can be shown that
\begin{equation}
\omega_1=\omega_2=\nu_0 e^{-\left|\Delta G_{12}\right|/k_B T}
\label{betrag}
\end{equation}
with the attempt frequency being given by $\nu_0 \approx D_0/l^2$
for $\left|\Delta G_{12}\right|\lesssim k_{B}T$ and $\nu_0 \approx
\pi \left|\Delta G_{12}\right| D_0/\left(k_B T l^2\right)$ for
$\left|\Delta G_{12}\right|\gg k_{B}T$.

Using Eqs.~\ref{p1} to \ref{betrag} we arrive at the final formula
for the diffusion constant for the case $\Delta G_{12}\ge 0$ (i.e.
$f \le 1$):
\begin{equation}
D_{>}=\frac{\nu _{0}}{1+K+f}\left( f+f^{2}\right) l^{2}
\label{Dlarger}
\end{equation}
In the opposite case, $\Delta G_{12} \le 0$ (i.e. $f \ge 1$), we
find
\begin{equation}
D_{<}=\frac{\nu _{0}}{1+K+f}\left( f^{-1}+1\right) l^{2}
\label{Dsmaller}
\end{equation}

Let us now consider special cases:

(\textit{i}) homogeneous DNA bendability, no ligands ($\Delta
G_{12}=0$, $\left[ L\right] =0$): In that case $f=1$, $K=0$. Both
formulas, Eqs.~\ref{Dlarger} and \ref{Dsmaller}, reduce to
$D_{>}=D_{<}=D_0$.

(\textit{ii}) homogeneous DNA bendability but ligands present
($\Delta G_{12}=0$, $\left[ L\right] >0$), cf.~Fig.~1(a): $f=1$
leads to
\begin{equation}
D_{>}=D_{<}=D=\frac{2D_{0}}{2+K}  \label{diff2}
\end{equation}

(\textit{iii}) rotational positioning sequence, no ligands present
($\left|\Delta G_{12}\right|\gg k_B T$, $\left[ L\right] =0$):
Equations \ref{Dlarger} and \ref{Dsmaller} reduce to
Eq.~\ref{deff} with $\left|\Delta G_{12}\right|=A \gg k_B T$, the
case that has been already discussed before \cite{kulic03b}.

(\textit{iv}) rotational positioning sequence, ligands present
with binding site exposed in the preferred orientational frame
($\Delta G_{12}\gg k_B T$, $\left[ L\right] >0$), cf.~Fig.~1(b):
Using $f\ll 1$ we obtain from Eq. \ref{Dlarger}
\begin{equation}
D_{>}=\frac{\pi \left|\Delta G_{12}\right|f}{k_B T}\frac{D_0}{1+K}
\label{diff4}
\end{equation}

(\textit{v}) rotational positioning sequence, ligands present with
binding site occluded in the preferred orientational frame
($\Delta G_{12}\ll k_B T$, $\left[ L\right] >0$), cf.~Fig.~1(c):
Here $f\gg 1$ and from Eq.~\ref{Dsmaller} we find
\begin{equation}
D_{<}=\frac{\pi \left|\Delta G_{12}\right|}{k_B T}\frac{D_0}{f+K}
\label{diff5}
\end{equation}

We are now in the position to check how effectively ligands reduce
repositioning in the various cases. We estimate in the following
the typical equilibration time on a 216 bp long template (as it
has been used in Ref.~\cite{gottesfeld02}) to be
$T_{70bp}=\left(216-146\right)^{2}bp^2/\left(2D\right)$. Let us
start with case (\textit{i}) where $D=D_0 \approx 580bp^{2}/s$.
This leads to the typical time $T_{70bp}=4s$. Adding now a ligand
with $[L]=100nM$ and $K_d=1nM$ (case (\textit{ii})) this leads to
a 50 fold reduction of the diffusion constant, $D\approx
12bp^2/s$, cf. Eq.~\ref{diff2}, and to an equilibration time
$T_{70bp}\approx 3.5 min$. If one uses a positioning sequence
instead with $\left|\Delta G_{12}\right|=9k_B T$ one finds in the
absence of ligands (case (\textit{iii})) from Eq.~\ref{deff}
$D\approx 2bp^2/s$ and $T_{70bp}\approx20 min$. Repositioning
experiments on such sequences are thus typically performed on time
scale of an hour to ensure equilibration
\cite{pennings91,gottesfeld02}. Adding now a ligand with
$[L]=100nM$ and $K_d=1nM$ and having its binding site facing the
solution in the preferred rotational frame (case (\textit{iv}),
Fig.~1(b)) we predict from Eq.~\ref{diff4} an additional dramatic
reduction of the diffusion constant by a factor of 100:
$D_{>}\approx 2\times 10^{-2} bp^2/s$ and $T_{70bp}\approx 34 h$.
In other words, in this situation one does not observe any
repositioning of the nucleosomes on the time scale of an hour.
This is in accordance with the experimental observations, cf.
Fig.~5, lane 1 and 4 in the study by Gottesfeld {\it et al.}
\cite{gottesfeld02}. On the other hand, for the case of a ligand
with same affinity and concentration but with the binding site in
the unfavorable orientation (case (\textit{v}), Fig.~1(c)) one
finds hardly any effect; in fact the diffusion constant as
compared to the ligand free case, case(\textit{iii}), is reduced
by approximately 1 percent, cf. Eq.~\ref{diff5}. In the experiment
\cite{gottesfeld02} these two cases were indeed indistinguishable
as seen in Fig.~5, lane 0, 2 and 3 in that paper.

\section{Transcription Induced Sliding}
\label{section3}

Gottesfeld et al.~\cite{gottesfeld02} also studied how nucleosomes
affect transcription. For that purpose the 216 bp DNA fragment
contained a T7 promoter in addition to the 5S positioning element.
The transcription reaction of the {\it naked} 216 bp fragment with
T7 RNA polymerase produced the 199 bp full-length RNA transcript.
Importantly this reaction was not affected by the presence of any
of the ligands. Also the nucleosome templates produced full length
transcripts with a very high yield, indicating that the RNA
polymerase was able to overcome the nucleosomal barrier. This was
also the case in the presence of ligands 2 and 3 whose binding
site face the octamer in the preferred rotational frame.
Remarkably the addition of ligand 1 or 4 blocked the
transcription. In fact, single round transcription assays showed
that the polymerase got stuck just within the major nucleosome
position. Moreover, an inspection of the nucleosome positions
showed that in the absence of any ligand or in the presence of
ligand 2 or 3 nucleosome repositioning took place. In other words,
transcription did not result in a loss of the nucleosome but in
its repositioning instead.

In the previous section we have shown that nucleosomes in the
presence of ligands 1 or 4 show a dramatic reduction of their
diffusion constant, cf. Eq.~\ref{diff4}. The Einstein relation
$\mu = D/k_B T$ provides a link between nucleosomal mobility $\mu$
and diffusion constant $D$ -- in case of thermodynamic
equilibrium. It is tempting to speculate that it is this
difference in nucleosomal mobility that is responsible for the
different outcome of the transcription experiment described in
Ref.~\cite{gottesfeld02}.

Let us first consider the case of a long DNA template with a
nucleosome positioned far from any of the DNA termini. Suppose
that an elongating RNA polymerase encounters such a nucleosome. If
the mobility of the nucleosome is large enough the RNA polymerase
would be able to push the nucleosome in front of it -- by pulling
the DNA in corkscrew fashion. In the most simple mean-field-type
approach the nucleosome will begin to slide with a constant speed
$v$ as a result of the imposed external load $F$ as follows:
\begin{equation}
v\left(F\right) = \mu F \label{mob}
\end{equation}

The polymerase slows down as a result of the force that it has to
exert on the nucleosome. According to Wang et al.~\cite{wang98}
(cf. also related studies \cite{juelicher98,wang98b}) the
force-velocity relation of RNA polymerase has typically the
following functional form
\begin{equation}
v\left(F\right) = \frac{v_0}{1+a^{\left(F/F_{1/2}\right)-1}}
\label{RNAP1}
\end{equation}
where $v_0$ is the velocity of the elongating complex in the
absence of an external load and $F_{1/2}$ is the load at which the
speed of the RNA polymerase is reduced to $v_0 /2$. $a$ is a
dimensionless fit parameter.

\begin{figure}
\begin{center}
\includegraphics*[width=8cm]{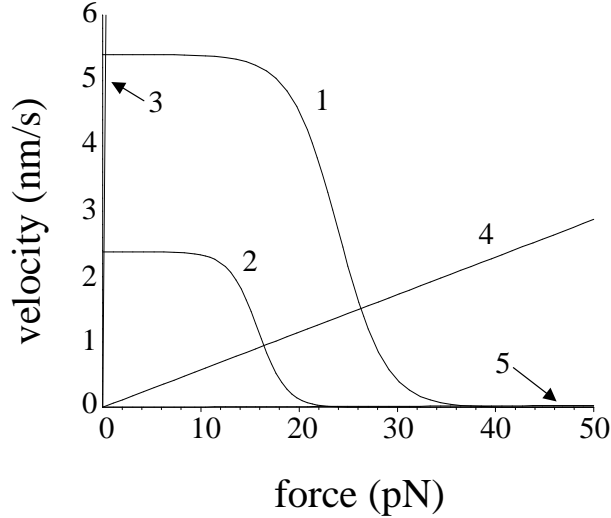}
\end{center}
\caption{Force-velocity relations. Curve "1" and "2" show the
relation between transcription-velocity and externally applied
load of RNA polymerase in two different cases (see main text).
Lines "3" to "5" give the force-velocity relation for nucleosomes
under an externally imposed force, again for three different
cases, cf. main text.}
\end{figure}

Equating Eqs.~\ref{mob} and \ref{RNAP1} we can determine the
average speed of an RNA polymerase that pushes a nucleosome in
front of it. The solution is found graphically in Fig.~2 by
determining the point of intersection between the corresponding
curves. Curves "1" and "2" in Fig.~2 give the force-velocity
relation of RNA polymerase, Eq.~\ref{RNAP1}, at two different
concentrations of pyrophosphate ($PP_i$), namely curve "1" is for
$1 \mu M$ $PP_i$ where $a=2 \times 10^4$, $F_{1/2}=24pN$ and $v_0
= 16 bp/s$ and curve "2" corresponds to $1 mM$ $PP_i$ for which
$a=5 \times 10^4$, $F_{1/2}=16pN$ and $v_0 = 7 bp/s$. In both
cases there is 1 mM nucleoside triphosphates (NTPs). Note that
these numbers give a good fit to the data of Wang et
al.~\cite{wang98} for the case of {\it Escherichia Coli} RNA
polymerase. As mentioned above in the experiment of Gottesfeld et
al.~\cite{gottesfeld02} a T7 RNA polymerase has been used and the
concentration of NTPs was $250-500 \mu M$. This means that curves
"1" and "2" can only be considered as rough estimates for the
force-velocity characteristics of the T7 RNA polymerase. The other
curves, "3" to "5", give the force-velocity relation,
Eq.~\ref{mob}, for the nucleosomes in various cases. Curve "3"
corresponds to the case when a nucleosome slides along an
isotropic DNA segment in the absence of any ligands (case
(\textit{i}) in the previous section). Curve "4" represents
corkscrew sliding along an anisotropic DNA with a barrier height
$9 k_B T$ as it is the case for the 5S positioning sequence (case
(\textit{iii})). And finally, curve "5" corresponds to the case
where in addition to such an anisotropic bendability the mobility
is slowed down by the presence of $100nM$ ligands with the ligand
binding site facing the solution in the preferred DNA bending
direction, case (\textit{iv}).

By inspecting the points of intersection between the curves we
come to the conclusion that RNA polymerase would be hardly slowed
down by the presence of a nucleosome on a homogenous track of DNA,
cf. point of intersection between line "3" with curve "1" (or "2")
in Fig.~2. We expect that the polymerase would easily push the
nucleosome in front of it without being slowed down. On the other
hand, the 5S positioning element should affect the transcription
rate by a considerable amount (cf. line "4" and curve "1" and
"2"); still the RNA polymerase might be able to push the
nucleosome ahead of it. Finally, in the case of added ligands the
nucleosome blocks the way of the nucleosome: the point of
intersection between curve "5" and "1" (or "2") is close to a
vanishing transcription velocity.

In the experiment \cite{gottesfeld02} there is, however, an
additional complication: The nucleosome is positioned at the
3'-end of the template. That means as soon as the polymerase
encounters the nucleosome (here after it has transcribed the first
$\approx 54$ bp) it would have to push the nucleosome {\it off}
the DNA template. What is the energetic cost of this process?
There are 14 binding sites between the DNA and the octamer, with a
10 bp distance between neighboring ones. It can be estimated
\cite{schiessel03} that the detachment of any of these 14
nucleosomal binding sites costs $\approx 6 k_B T$. However, the
overall energetic cost of undressing the nucleosome is smaller:
When pulling 10 bp off the octamer one binding site is opened but
10 bp are released on the other side that gain roughly $4 k_B T$
elastic energy by going from the wrapped, bent state to the
straight state. In total, a shift of the DNA by 10 bp cost
therefore only $2 k_B T$ which corresponds to a force of just 2
pN. This additional force can be easily supplied by the RNA
polymerase.

Therefore our calculation leads to the prediction of the following
effect of the RNA polymerase on the nucleosome: (1) In the
ligand-free case the RNA polymerase is able to produce the
full-length transcript pushing the nucleosome off the template.
(2) If a ligand is bound to the nuclesomal DNA the nucleosome is
immobile and the polymerase stalls as soon as it encounters the
nucleosome. Whereas the second prediction is indeed in agreement
with the experimental observations, the first one is not. In that
case transcription does not lead to the loss of the nucleosome but
instead to its repositioning on the template \cite{gottesfeld02}.
The experimental findings even indicate that the nucleosome -- as
a result of the transcription -- is effectively moving upstream!
In fact, such effects have been studied in detail before and let
to proposition of a spooling mechanism
\cite{studitsky94,studitsky95,studitsky97,bednar99}; we will here,
however, delegate a discussion of these experiments and their
interpretation to our discussion section.

\begin{figure}
\begin{center}
\includegraphics*[width=8cm]{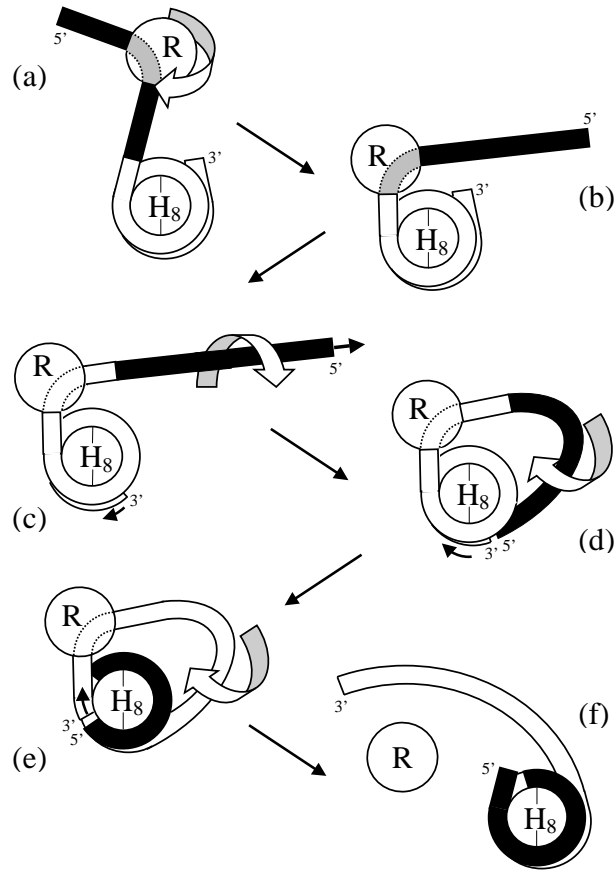}
\end{center}
\caption{Tentative model for nucleosome repositioning via an
extranucleosomal loop: The transcribing polymerase encounters in
(b) the nucleosome. It either gets stuck (if the nucleosome is
immobile) or (c) it starts to pull the DNA in a corkscrew fashion
from the nucleosome "undressing" it at the other end. (d) The free
DNA end adsorbs on the nucleosomal binding sites that have just
been exposed. As a result an {\it extranucleosomal} loop has
formed. (e) The RNA polymerase continues to pull the DNA around.
(f) Finally the other DNA end is released. As a result of the
transcription the nucleosome has been transferred to the other
(former free) end of the DNA.}
\end{figure}

In order to explain the experimental observations of
Ref.~\cite{gottesfeld02} we propose a new mechanism that is
depicted in Fig.~3. (a) At the beginning of the transcription (the
first 54 bp in Ref.~\cite{gottesfeld02}) the RNA polymerase walks
along the free DNA section (shown in black) in a corkscrew
fashion. (b) The polymerase comes into contact with the
nucleosome. At this stage the polymerase gets stuck if the
nucleosome is immobile. (c) If the nucleosome is mobile the
polymerase pulls on the DNA, undressing the nucleosome at the
other end (the 3' end). During this process the polymerase and the
octamer (H$_8$) are not moving with respect to each other and it
is only the DNA that is performing a corkscrew motion. (d) After
enough nucleosomal contact points (at the 3' end) are exposed to
the solvent the 5' end might adsorb on these contact points,
forming an {\it extranucleosomal} loop. (e) The DNA continues to
circle around the polymerase-nucleosome complex via the corkscrew
mechanism. (f) When the 3' end reaches the polymerase this end is
released from the nucleosome. As a result one has again an
end-positioned nucleosome but now it is the promoter end that is
wrapped on the nucleosome. A section of the original positioning
sequence (shown in white) forms now the free tail.

\begin{figure}
\begin{center}
\includegraphics*[width=3cm]{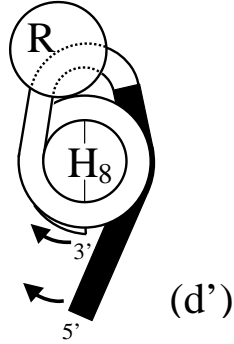}
\end{center}
\caption{Alternative version of the extranucleosomal loop model:
In stage (d) of Fig.~3 the bend induced by the RNA polymerase
leads to the formation of a very small extranucleosomal loop. The
5' end forms then a tail on the nucleosome.}
\end{figure}

This mechanism always transfers the nucleosome from one end of the
DNA template to the other. In principle, it is also possible that
a smaller loop forms with the 5' end forming an overhanging tail,
cf. Fig.~4. Such a small loop might be possible since the RNA
polymerase induces a bend on the DNA. The RNA polymerase will then
again pull the DNA around via the corkscrew mechanism. Due to the
presence of the loop the 5' tail might only be able to adsorb
beyond the dyad after the 3' end is released. At this point the
nucleosome has effectively made a step upstream. The step length
is the sum of the length stored in the loop plus the number of bp
of the 3' end that were still adsorbed at the point of its
release. It is possible that the 3' is released at a point where
it was still associated with a few binding sites (each binding
site just contributes on the order of $2 k_B T$). The typical
upstream step length is then a few tens of bp. An interesting
feature of this variant of the model is that the step length
should not depend on the length of the originally free DNA portion
(shown in black in Fig.~4). In other words: If the nucleosome was
initially positioned at one end of the template (due to some
positioning sequence), after transcription it is shifted upstream
to a new position by a distance that is independent of the length
of the DNA template.

\section{Discussion}
\label{section4}

We provide here a critical discussion of our results, mainly
focusing on alternative mechanisms that allow RNA polymerase to
get around nucleosomes. Before doing so we mention an important
assumption on which our whole analysis is based: Bound ligands do
not act {\it across} the two turns. If the ligand would somehow
bind to both turns of the nucleosome, it could effectively block
any nucleosome dynamics, whether it is induced by bulge or twist
defects. In fact, such a situation occurs in the presence of
linker histones H1 or H5 that bind the in- and outgoing DNA
together in a stem-like region. It has been observed that there is
no nucleosome repositioning in this case \cite{pennings94}.
However, the cocrystal structure of a nucleosome core particle
with bound ligands \cite{suto03} indicates that ligands bind
locally to one turn without affecting the other one. This supports
our assumption -- namely that a ligand blocks twist defects only.

This brings us to the discussion of polymerase-induced nucleosome
repositioning. The experiment by Gottesfeld et
al.~\cite{gottesfeld02} showed that nucleosomes survive
transcription but it was not possible to deduce from the data
whether transcription through a nucleosome leads to its
repositioning along DNA. There is, however, a long series of
experiments that have focused on this point
\cite{studitsky94,studitsky95,studitsky97,bednar99}. Also in these
experiments a bacteriophage RNA polymerase has been used, namely
that of SP6. The standard 227 bp template includes an SP6 promoter
and a nucleosome positioning sequence \cite{studitsky94}.
Typically the nucleosome is positioned at the promoter distant
end. Transcription results in an upstream displacement to the
other end, i.e. by 80 bp \cite{studitsky94}. Whether this step
length reflects a built-in step length of the repositioning
process or whether the nucleosome is displaced from one end to the
other has been checked by adding an extra length to the DNA
template at either end. Adding extra 50 bp at the promoter side
(the 5' end) the upstream step is typically 90 bp, i.e., it does
not increase much. This might indeed indicate that the
displacement process has a natural 80 to 90 bp step length. On the
other hand, addition of 35 bp to the 3' end has surprisingly also
an effect on the upstream step length that shows now three smaller
values, namely 40, 60 and 75 bp \cite{studitsky94}. Finally, going
to a much larger template by adding 126 bp at the promoter end led
to another surprise: In this case the nucleosome is transferred
from one end to the other as a result of the transcription
\cite{studitsky95}.

How can these observations be rationalized? Studitsky et
al.~\cite{studitsky94} introduced the "spooling" mechanism, cf.
their Fig.~7: As the polymerase encounters the nucleosome it
continues to transcribe by prying off the DNA from the octamer.
After the polymerase has proceeded far enough into the nucleosomal
DNA the DNA behind might attach to the now exposed nucleosomal
binding sites. This results in an {\it intra}nucleosomal loop. The
polymerase travels around the nucleosome inside this loop. When
reaching the other end the loop disappears and as a result the
nucleosome steps upstream by the extra DNA length that has been
stored in that loop. The step lengths observed in the experiments
have then to be interpreted as the loop sizes. A preferred value
would then be around 80 bp. Studitsky et al. explained the much
shorter step lengths observed in the case of a template with a DNA
extension on the promoter distant site as a result of "octamer
slippage" before the spooling mechanism comes into play with the
usual 80 bp upstream step. Finally, the end to end transfer on the
long 353 bp template indicates a large loop that stores 180 to 200
bp \cite{studitsky95}.

In fact, these observations and their explanation are entirely
consistent. One should nevertheless ask whether our {\it
extra}nucleosomal loop model provides also a picture consistent
with these experimental facts. In fact, the model depicted in
Fig.~3 predicts an end-to-end transfer of the nucleosome as it has
been observed for the longest template discussed above. The
modified model with a small extranucleosomal loop as depicted in
Fig.~4 leads to a smaller upstream step of the octamer whose value
depends on microscopical details but should be on the order of a
few tens of bp. So this picture could in fact also explain the
typical 80 bp shifts observed in several cases. This leads us to
the surprising conclusion that either mechanism, the extra- and
the intranucleosomal one, is consistent with the observations. It
is only the smaller steps where Studitsky et al. suggested octamer
slippage to occur that might ironically speak in favor of their
model. When the nucleosome steps back by 40 bp it might have first
slid 35 bp to the 3' end and then go back by 80 bp with either
mechanism. However, the fact that after transcription some
nucleosomes were found 60 and 75 bp upstream might support the
intranucleosomal loop picture: First the nucleosome slides a short
distance (but not up to the DNA terminus) and then steps back by
80 bp in an intranucleosomal loop. Nevertheless it seems
impossible to exclude from these experimental observations one or
the other mechanism and it might well be the case that both play a
role.

Another feature that has been observed during the transcription
"through" nucleosomes is a characteristic pausing pattern of the
polymerase \cite{studitsky95,protacio96}. Studitsky et
al.~\cite{studitsky95} reported for their SP6 system pausing with
a 10 bp periodicity to occur up to the dyad where it disappears.
Also Protacio et al.~\cite{protacio96} find pausing with this
periodicity, however, also extending far beyond the dyad. The
ladder system uses T7 RNA polymerase and the 5S positioning
element as in Ref.~\cite{gottesfeld02}. Studitsky et al. interpret
their observations with their spooling model: Once the loop has
formed the polymerase might not be able to continue with
elongating since it would have to corkscrew through the loop and
this process might be too costly if not even sterically forbidden.
Instead pausing occurs up to the point when the loop reopens
through a spontaneous fluctuation. The loop formation (and the
concomitant pausing) might happen with a 10 bp periodicity since
the bend induced by the polymerase might help the loop formation
every 10 bp. Once the dyad has been reached the last loop forms
that is finally broken {\it ahead} of the polymerase, allowing the
polymerase to transcribe from now on without interference from the
octamer. Further support for this idea was given by removal of DNA
behind elongating complexes that have been arrested just at the
nucleosomal border. Resuming transcription the polymerase was able
to elongate into the nucleosome much further without pausing
before it encountered a first pausing side. This was interpreted
again as a fact supporting the spooling model \cite{studitsky95}:
The formation of the loop was only possible when enough DNA was
available at the 5' end.

We believe that these observations are also consistent with the
extranucleosomal loop picture. The 10 bp pausing pattern might
reflect the 10 bp periodicity of the bending energy of the
positioning sequence. Enhanced pausing might occur once the loop
has formed due to an enhanced friction of the corkscrewing DNA.
And the disappearance of pausing sites beyond the dyad (which is
not for all situations the case, cf. Ref.~\cite{protacio96}) might
reflect the termination of an interaction between the polymerase
and DNA wrapped close to the dyad. In case of the 5' end forming a
tail, as shown in Fig.~4, this end might not be able to adsorb
beyond the dyad as long as the intranucleosomal loop is present so
that the friction or entanglement between the components decreases
once the polymerase passes the dyad.

This brings us to the next point of our discussion. One might
wonder whether such intra- or extranucleosomal loops could be
directly "seen" in electron micrographs. In fact, cryomicroscopy
has been performed for such complexes \cite{bednar99}.
Unfortunately, also here the situation is rather complex. When the
polymerasewas arrested after transcribing 23 bp into the
nucleosome the electron cryomicrographs showed complexes with one
DNA tail. The length of that tail was considerably longer than the
tail in the absence of RNA polymerase. This was interpreted as
being due to a polymerase-induced DNA unwrapping. Interestingly
our corkscrew sliding scenario also leads to a tail lengthening
without the necessity of DNA unpeeling, cf. Fig.~3(c). The
polymerase was also arrested further into the nucleosome (42 bp),
a location at which intra- or extranucleosomal loops should be
expected. Loops were, however, not observed (at least not large
ones), instead there was a considerable fraction of two-tailed
intermediate states. These closed transcription intermediates were
interpreted as states that resulted from the collapse of an
internucleosomal loop, cf. Fig.~7 in Ref.~\cite{bednar99}. In our
opinion such an explanation (being an attempt to reconcile the
spooling model with the two-tail intermediates) is not obvious,
even though this picture cannot be excluded. On the other hand,
when the polymerase is stalled after a small extranucleosomal loop
has formed, two-tail intermediates should in fact be expected. In
Fig.~4 the 5' end is forming the only tail. But it is also
possible that the 3' desorbs up to the dyad where the loop blocks
further unpeeling. This leads to two-tail complexes where both
ends form tails of varying lengths.

In conclusion we agree that the experiments of Studitsky et
al.~\cite{studitsky94,studitsky95,studitsky97,bednar99} are indeed
compatible to their spooling model. However, we have shown that
also our extranucleosomal loop mechanism gives a consistent
explanation of these experiments. If the recent experiments by
Gottesfeld et al. \cite{gottesfeld02} had been performed for the
same system under identical conditions, then their observation of
transcription blockage via ligands would speak in favor of our
model. As it stands it is not clear which mechanism is responsible
for transcription through nucleosomes in the various cases.

We should mention that the two different scenarios involving
intra- and extranucleosomal loops lead to dramatically different
pictures for transcription on multinucleosomal templates. Whereas
the elongating RNA polymerase could easily get around all the
nucleosomes via intranucleosomal loops, our extranuclesomal
variant relies on the finite length of the DNA. This mechanism
would cease to work for the multinucleosomal situation. In fact,
transcription on reconstituted multinucleosomal templates showed
that T7 RNA polymerase is under certain conditions capable of
disrupting completely the nucleosomal cores
\cite{heggeler95,heggeler00}. Electron micrographs show the
transcribed section to be freed of nucleosomes and parts of the
histones being transferred to the nascent RNA chain
\cite{heggeler00}. Interestingly upon addition of some nuclear
extract the nucleosomal template seem to survive during
transcription \cite{heggeler95}. This shows that the {\it in vivo}
situation might be rather complex involving additional factors
mediating between polymerase and nucleosomes.

\section{Acknowledgements} We are thankful to Karolin Luger for
sharing experimental results \cite{suto03} prior to publication.
We thank R. Bruinsma and R. Golestanian for discussions. F.M.
acknowledges the hospitality of the MPIP in Mainz where this work
was initiated.



\end{document}